%% file: Rauch.tex
\documentclass[cits]{PoS}

\usepackage{amsmath}
\usepackage[utf8]{inputenc}

\DeclareMathOperator{\Tr}{Tr}

\title{Multi-boson Production in Weak Boson Fusion}

\ShortTitle{Multi-boson Production in Weak Boson Fusion}

\author{\speaker{Michael Rauch}\\
        Institute for Theoretical Physics, Karlsruhe Institute of Technology (KIT), Germany\\
        E-mail: \email{michael.rauch@kit.edu}}

\author{Francisco Campanario\\
        Theory Division, IFIC, University of Valencia-CSIC, E-46980 Paterna, Valencia, Spain\\
        E-mail: \email{francisco.campanario@ific.uv.es}}

\author{Bastian Feigl\\
        Institute for Theoretical Physics, Karlsruhe Institute of Technology (KIT), Germany}

\author{Oliver Schlimpert\\
        Institute for Theoretical Physics, Karlsruhe Institute of Technology (KIT), Germany}

\abstract{%
The production of multiple gauge bosons via weak boson fusion is an
important process at the LHC. It is relevant as a background process
appearing in many searches and measurements, but also serves as a signal
process when studying new-physics contributions to triple and in
particular quartic gauge boson vertices.

We first review the theoretical status of multi-boson production in weak
boson fusion and present the current state of the art. In the second
part, the impact of anomalous gauge couplings on this class of processes
is discussed.
}

\FullConference{XXII. International Workshop on Deep-Inelastic Scattering and Related Subjects,\\
		28 April - 2 May 2014\\
		Warsaw, Poland}

\begin{document}

\section{Introduction}
\label{sec:intro}

The production of multiple bosons in weak-boson fusion, often also
called vector-boson fusion (VBF) or vector-boson scattering (VBS), yields
a distinct signature, which is characteristic for the
whole class of weak-boson fusion processes. It consists of two jets in
the forward region of the detector, the so-called tagging jets, and a
reduced jet activity in the central region. The leptonic decay products
of the final-state bosons are typically in rapidity between the tagging
jets.  This process class can also be seen as a two-sided deep-inelastic
scattering (DIS) process, where the electron side of the DIS process is
replaced by the other quark in each case.

The process class of weak-boson-fusion has first been studied in the
context of Higgs
searches~\cite{hep-ph/9412276%
}, where the two tagging jets serve
as an experimental trigger and allow for a reduction of background
processes. Later on, the focus has been extended to VBF production of
single gauge bosons ($W$, $Z$, $\gamma$)~\cite{hep-ph/9605444%
} and finally to diboson
production~\cite{hep-ph/0603177,hep-ph/0604200,hep-ph/0701105,0907.0580,1006.4237,1209.2389,1212.5581,1309.7259}.
In this overview, we will restrict ourselves to the production of a pair
of gauge bosons, since the Higgs-pair production process is discussed in a
separate article~\cite{baglio}. These studies include the leptonic decays of
the gauge bosons as well as off-shell and non-resonant contributions.

\begin{figure}
{\centering\includegraphics[width=0.8\textwidth]{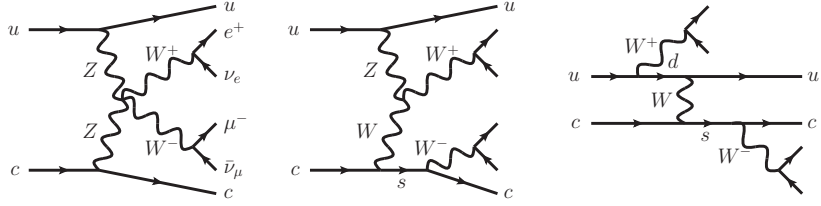}\\}
\caption{Example Feynman diagrams for $W^+W^-$ production via VBF
including leptonic decays of the gauge bosons.}
\label{fig:feynman}
\end{figure}
A set of typical Feynman diagrams is shown in Fig.~\ref{fig:feynman},
taking $W^+ W^-$ production via VBF as an example process.  The
final-state $W$ can either be attached directly to the quark lines, as
in the diagram on the right-hand side, or one or both are attached to
the intermediate t-channel gauge boson. This gives then rise to triple
and quartic gauge boson vertices, as shown in the middle or left-hand
Feynman diagrams, respectively. Therefore, the
non-Abelian structure of the weak interaction can be studied with these
processes. Furthermore, they give
access to triple and, in particular, quartic anomalous gauge
couplings~\cite{Buchmuller:1985jz%
,1309.7890}. 

Furthermore, these processes form a background for many collider
searches, both within the Standard Model and beyond. For example, they
constitute an irreducible background to VBF production of Higgs bosons
where the Higgs decays into gauge bosons. They are important both for
the discovered light Higgs state with a mass of around 125~GeV, but also
for searches for heavy Higgs bosons. There, a possibly large width of
the heavy Higgs can lead to significant interference effects between the
Higgs signal and the continuum background diagrams~\cite{1307.1347}.

On the experimental side, evidence for VBF production of the single
gauge boson process $Zjj$~\cite{1401.7610} and the diboson process
$W^\pm W^\pm jj$~\cite{1405.6241} has been established so far by the
ATLAS experiment. The latter process has also been studied by
CMS~\cite{CMS-PAS-SMP-13-015}, where the observed significance does not
yet reach the 3$\sigma$ level.


\section{Higher-order corrections and parton-shower effects}
\label{sec:nlo}

Next-to-leading-order (NLO) QCD corrections for diboson production via
VBF have been calculated in
Refs.~\cite{hep-ph/0603177,hep-ph/0604200,hep-ph/0701105,0907.0580,1209.2389,1309.7259}.
Most of these calculations have been performed in the so-called VBF
approximation.  There, contributions from s-channel diagrams, which can
also be viewed as triboson production processes where one of the gauge
bosons decays hadronically, as well as effects from interference between
$t$- and $u$-channel diagrams are neglected.
This approximation has been checked explicitly at leading order (LO) by
performing the full calculation of electroweak $W^+W^+jj$ production in
Ref.~\cite{1209.2389}, and extending this comparison also to the
QCD-induced production process in Ref.~\cite{1311.6738}.
When applying VBF cuts, which exploit the typical structure of these
processes, 
$m_{j_1 j_2} > 600 \text{ GeV}$, 
$y_{j_1} \times y_{j_2} <0 $,
$|y_{j_1} - y_{j_2}| > 4 $ and $y_{j_{\min}} < y_{\ell} < y_{j_{\max}}$, 
where $j_1$ and $j_2$ denote the first and second hardest jet, the
difference between the full calculation and the VBF-approximated one
becomes negligible. This set of cuts also allows for an efficient
suppression of the corresponding QCD-induced processes, which have been
studied in much detail in
Refs.~\cite{1311.6738,1007.5313%
}.

\begin{figure}
{\centering
\includegraphics[width=0.7\textwidth]{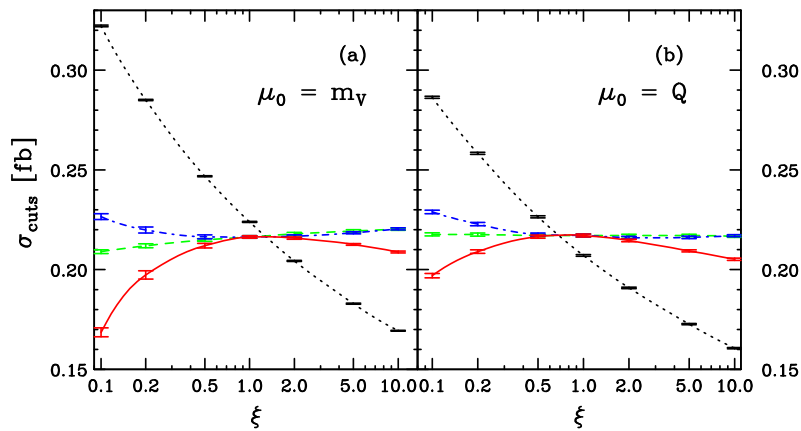}\\}
\caption{Dependence of the VBF-$W^+Zjj$ production cross sections on the
scale choice $\mu=\xi\mu_0$. Results are shown for LO (dotted black),
which only depends on the factorization scale, and for NLO when varying
the factorization scale only (dot-dashed blue), the renormalization
scale only (dashed green), or both jointly (solid red).
\textit{Left:} Fixed scale $\mu_0=m_V=\frac{m_W+m_Z}2$, \textit{right:}
momentum transfer $\mu_0=Q$.  Figure from
Ref.~\cite{hep-ph/0701105}.}
\label{fig:WZscale}
\end{figure}
The dependence on renormalization and factorization scale at LO and NLO
is shown exemplarily for $W^+Zjj$ VBF-production in Fig.~\ref{fig:WZscale}.
The LO cross section shows a sizable scale dependence of $\pm 10\%$ when
varying the value between $\frac12$ and $2$ times the central value. At
NLO this is strongly reduced to a value of $\pm 2\%$ for integrated
cross sections and up to $\pm 6\%$ in distributions. Thereby, taking the
dynamic scale $Q$, defined as the momentum transfer of the exchanged
vector boson in VBF diagrams, for each quark line, proves advantageous.
As one can see in the figure, both at LO and NLO the scale variation is
reduced compared to a fixed scale. The $K$ factor is close to
unity for both scale choices. In differential distributions, the NLO QCD
corrections can lead to relevant changes in shape. Therefore, a simple
rescaling with the integrated K factor is not sufficient. Also here,
choosing the momentum transfer as a scale leads to flatter differential
K factors than a fixed scale. A dedicated Monte Carlo tool to calculate
these processes at NLO QCD at a fully differential level is available
with the program VBFNLO~\cite{0811.4559%
}.

The next step is then to interface these NLO calculations with parton
shower, which combines the advantages of both approaches. The normalization of
the cross section is correct to NLO and the emission of the extra,
third, jet is accurately described also when it possesses large
transverse momentum. At the same time, the Sudakov suppression at small
transverse momentum is correctly modeled by the parton shower, and it
is possible to generate events not only at the parton, but also at the
hadron level.
Results have been presented for $W^+W^+jj$~\cite{1108.0864},
$W^+W^-jj$~\cite{1301.1695} and $ZZjj$~\cite{1312.3252} production via
VBF. All calculations available so far make use of the POWHEG-BOX
tool~\cite{1002.2581}, which employs the POWHEG
formalism~\cite{hep-ph/0409146%
} to combine NLO 
calculations and parton shower.
 
\begin{figure}[t]
{\centering
\includegraphics[width=0.38\textwidth]{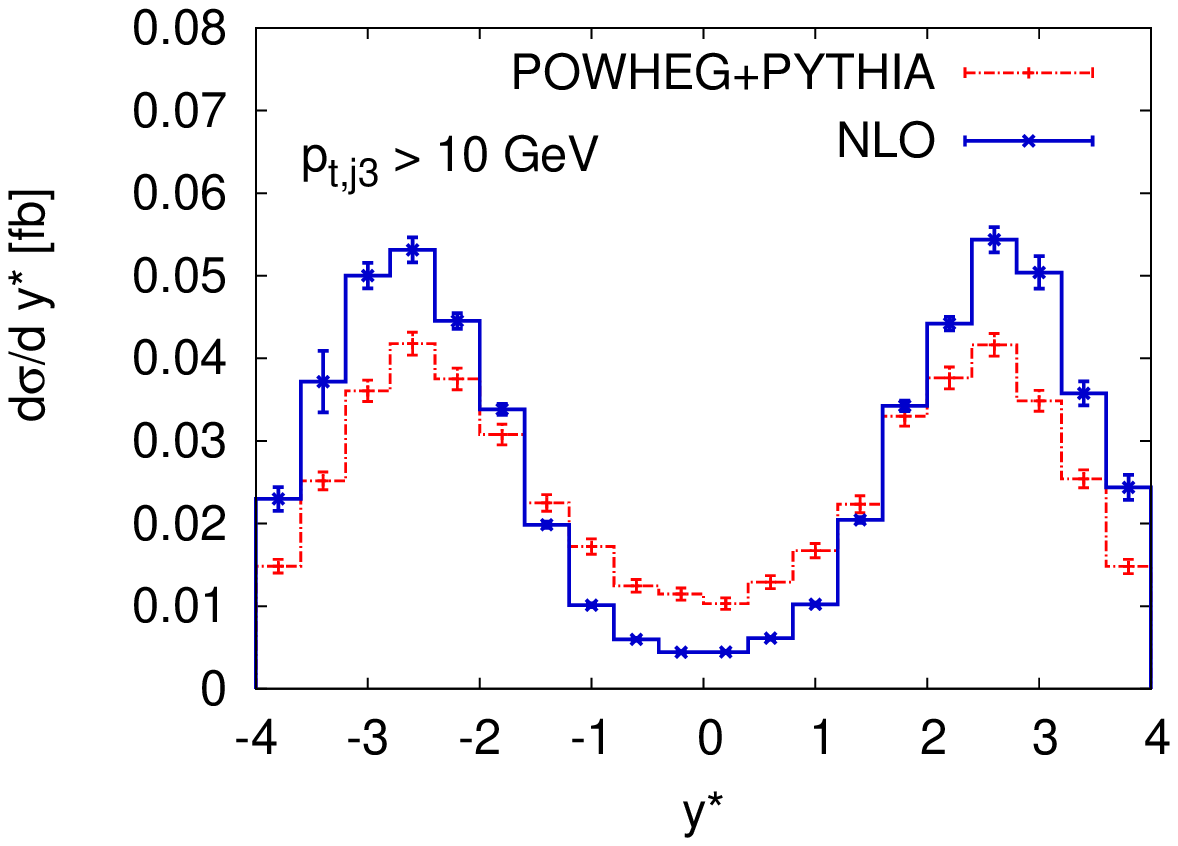}\qquad
\includegraphics[width=0.38\textwidth]{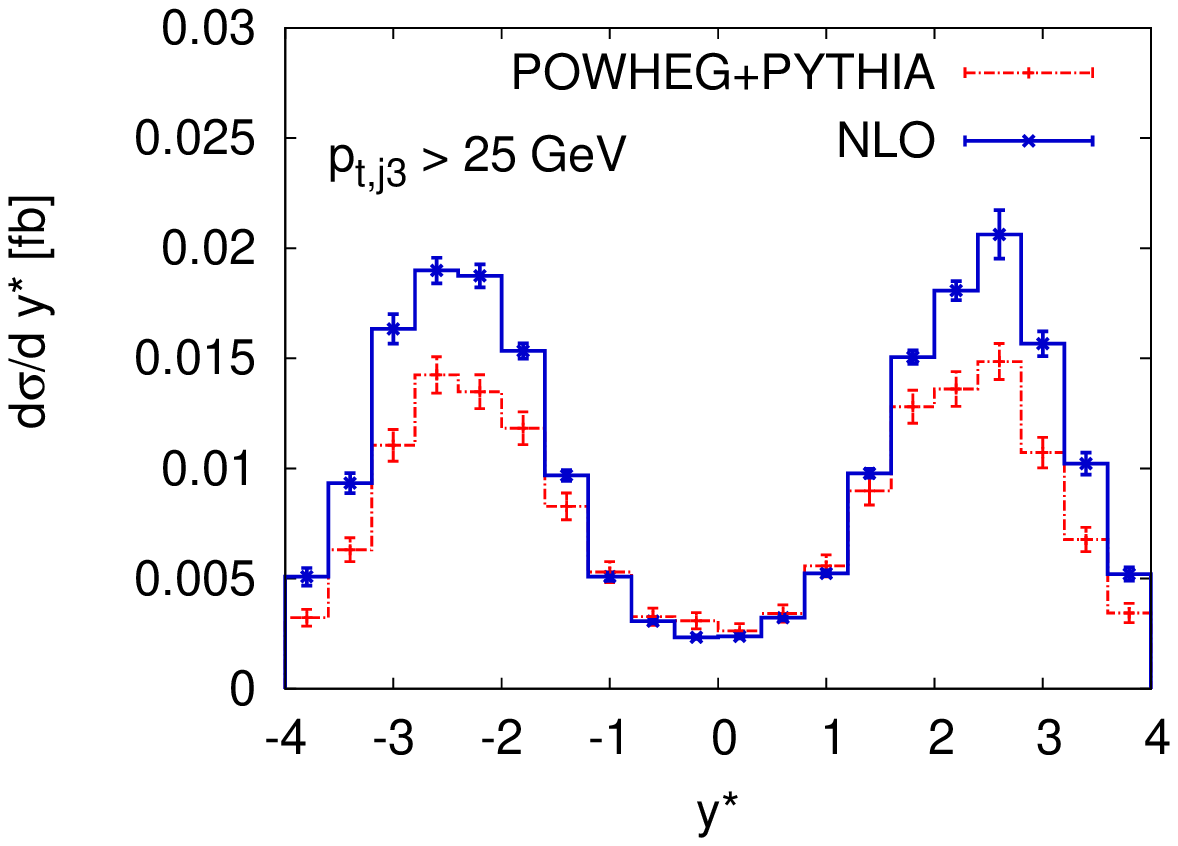}
\\}
\caption{Differential cross section of the process $W^+W^-jj$ for the
relative position of the third-hardest jet with respect to the two
tagging jets, $y^*$, for the pure NLO QCD calculation (solid blue) and
interfaced with parton shower (dot-dashed red). Shown are results for
two different values of the minimum transverse momentum of the third
jet.
Figure taken from Ref.~\cite{1301.1695}.}
\label{fig:WWystar}
\end{figure}
In Fig.~\ref{fig:WWystar} we show effects of the parton shower on the
emission of the extra jet for $W^+W^-jj$ production via VBF.
The kinematic variable studied is the relative position of the
third-hardest jet with respect to the two tagging jets, $y^* = y_{j_3} -
\frac{y_{j_1} + y_{j_2}}2$. Imposing only a loose cut on its transverse
momentum, $p_{T,j_3} > 10 \text{ GeV}$, displayed on the left-hand side,
we see that compared to the pure NLO calculation, the parton shower
predicts extra jet activity in the central rapidity region between the
two tagging jets. In contrast, the cross section for emission close to
one of the tagging jets is reduced. The extra emission in the central
region has mostly small transverse momentum, as can be seen from the
comparison with the plot on the right-hand side, where the minimum
$p_{T,j_3}$ has been increased to 25 GeV. There, the predictions with and
without parton shower agree in the central region. An important
technique in separating VBF processes from the corresponding QCD-induced
ones is applying a so-called mini-jet veto, i.e.\ requiring that no
extra emission above a certain, typically rather small, minimal
transverse momentum takes place in the region between the two tagging
jets. These results show that parton-shower effects can induce relevant
differences compared to a pure NLO QCD calculation.

\section{Anomalous gauge couplings}
\label{sec:anom}

The appearance of quartic gauge boson vertices makes diboson production
via weak-boson fusion an excellent process to test anomalous
contributions to these vertices. A helpful tool to describe the
deviations from the Standard Model (SM) prediction in a consistent,
Lorentz- and gauge-symmetry conserving way is the use of effective
field theory (EFT). There, new-physics contributions are assumed to be
too heavy to produce them directly, but their effects can modify the
interactions between the SM particles and appear as higher-dimensional
operators in the Lagrangian: $\mathcal{L}_{EFT} = \mathcal{L}_{SM} +
\sum_{d>4} \sum_{i} \frac{c_i}{\Lambda^{d-4}} \mathcal{O}_i$, where $d$
is the dimension of the operators and $\Lambda$ the scale of new
physics. The building blocks for the operators $\mathcal{O}_i$, which
are Lorentz scalars and gauge singlets, are the field strength tensors
of the gauge fields, the Higgs and fermion fields, and partial and
covariant derivatives. Below $\Lambda$, where this approach is valid,
only the operators with lowest dimensions can give large contributions
and should therefore be kept. For the VBS processes, these are the
operators of dimension~6 and dimension~8~\cite{Buchmuller:1985jz,1309.7890}. The former
can also be constrained by processes with triple gauge boson vertices only,
while the bosonic operators of the latter generate only vertices with at least
four bosons. Therefore they are particularly relevant for VBS.

If relevant scales of the process become larger than $\Lambda$, the EFT
approach breaks down. An indication for this can be the violation of
unitarity in the scattering amplitude. In experimental searches one has
to make sure that the sensitivity is not driven by phase-space regions
where unitarity is violated. Several possibilities for this task are
possible, one of the simplest being a step function which turns off
anomalous couplings for large energies. Other approaches just dampen
the additional contributions and give a smoother cut-off. Examples are
the K-matrix method~\cite{0806.4145}, where the amplitude is projected such
that its value is exactly at the unitarity bound, or a dipole form factor
$\left(1+\frac{m_{VV}}{\Lambda_{FF}^2}\right)^{-p}$, where the exponent $p$ is
an additional free variable. The last approach is used for the results in the
following. A dedicated tool to calculate the bound where unitarity violation
happens and a corresponding dipole form factor is available at
Ref.~\cite{VBFNLOformfac}. 

\begin{figure}[t]
{\centering
\includegraphics[width=0.45\textwidth]{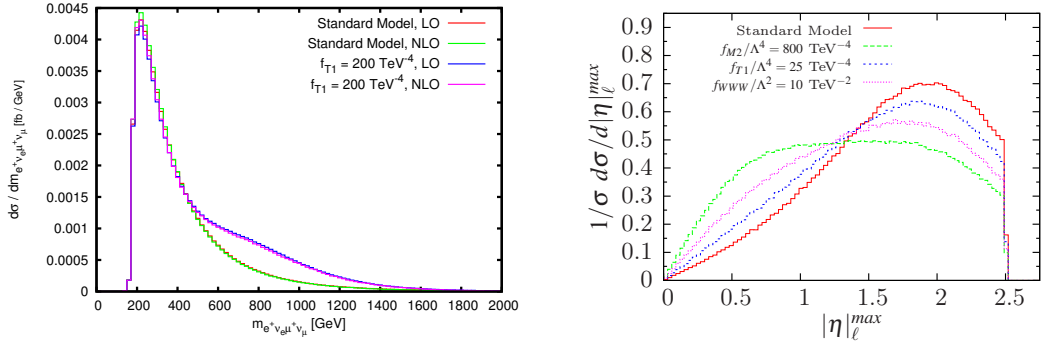}\qquad
\scalebox{0.85}{\input{kap5_etaLmax.tex}}
\\}
\caption{\textit{Left:}
Differential cross section for the invariant mass distribution of the
two lepton, two neutrino system in VBF-$W^+W^+jj$ production with and
without anomalous $T_1$ coupling at LO and NLO.
Figure taken from Ref.~\cite{1309.7890}.
\textit{Right:}
Normalized differential cross section for the pseudorapidity of the
leading lepton in VBF-$W^+W^-jj$ production for the SM and three
different operators of anomalous gauge couplings.
Figure taken from Ref.~\cite{schlimpert}.
}
\label{fig:anom}
\end{figure}
On the left-hand side of Fig.~\ref{fig:anom} we first show the influence
of anomalous couplings on the NLO QCD corrections for $W^+W^+$
production via VBF. As operator we choose $\mathcal{O}_{T_1} =
\left[W^{\alpha\nu}W_{\mu\beta}\right] \times
\left[W^{\mu\beta}W_{\alpha\nu}\right]$ as an example.
Starting at an invariant $W^+W^+$ mass of about 500 GeV, the additional
contributions due to the anomalous coupling lead to an increase of the
cross section, before the dipole form factor with
$\Lambda_{FF}=1188\text{ GeV}$ and $p=4$ damps the contribution again at
higher invariant masses. The effect of NLO corrections is small both
without and with anomalous couplings and does not influence the extraction
of anomalous couplings.
The right-hand side of Fig.~\ref{fig:anom} presents the normalized
differential cross section for the pseudorapidity of the leading lepton
in VBF-$W^+W^-jj$ production. Besides the SM, results are also shown for
the dimension-8 operators $\mathcal{O}_{T_1}$ and $\mathcal{O}_{M_2} =
\left[B^{\mu\nu}B_{\mu\nu}\right] \times \left[\left(D^\beta\Phi\right)^\dagger
\left(D_\beta\Phi\right)\right]$ as well as the dimension-6 operator
$\mathcal{O}_{WWW} =
\Tr\left[W_{\mu\nu}W^{\nu\rho}{W_{\rho}}^{\mu}\right]$. The anomalous
couplings always lead to a shift from larger to smaller
pseudorapidities, but the size of the shift and therefore the shape of
the distributions differs for the various operators. This feature could
eventually be used to distinguish different operators once a deviation
from the SM prediction has been established.

\section{Conclusions}
\label{sec:concl}

Diboson production via weak-boson fusion is an important process for the
LHC. It is a background to many searches and measurements, both within
and beyond the Standard Model. Due to the appearance of Feynman graphs
with quartic gauge boson vertices it can also be used as a signal
process in its own right and allows to study anomalous triple and in
particular quartic gauge couplings. NLO QCD corrections have been
calculated and turn out to be modest. Choosing the momentum transfer via
the t-channel gauge bosons as a scale thereby proves advantageous. For
a number of processes also a combination with parton shower is
available. These show enhanced rates in the central region for the
third-hardest jet at low transverse momentum. 
Experimental results for the VBF processes have started to appear
only very recently due to still rather small numbers of events collected
so far at the LHC. The prospects to study these processes in more detail
at the upcoming run of the LHC and other future hadron colliders are
very promising.

\acknowledgments
We would like to thank the organizers for the invitation and the
stimulating atmosphere at the conference. We would also like to thank
D.~Zeppenfeld for many helpful discussions.
MR acknowledges support by the Deutsche Forschungsgemeinschaft via the
Sonderforschungsbereich/Transregio SFB/TR-9 ``Computational Particle
Physics''.

\end{document}

%% file: kap5_etaLmax.tex
\begingroup
  \fontfamily{cmr}%
  \selectfont
  \makeatletter
  \providecommand\color[2][]{%
    \GenericError{(gnuplot) \space\space\space\@spaces}{%
      Package color not loaded in conjunction with
      terminal option `colourtext'%
    }{See the gnuplot documentation for explanation.%
    }{Either use 'blacktext' in gnuplot or load the package
      color.sty in LaTeX.}%
    \renewcommand\color[2][]{}%
  }%
  \providecommand\includegraphics[2][]{%
    \GenericError{(gnuplot) \space\space\space\@spaces}{%
      Package graphicx or graphics not loaded%
    }{See the gnuplot documentation for explanation.%
    }{The gnuplot epslatex terminal needs graphicx.sty or graphics.sty.}%
    \renewcommand\includegraphics[2][]{}%
  }%
  \providecommand\rotatebox[2]{#2}%
  \@ifundefined{ifGPcolor}{%
    \newif\ifGPcolor
    \GPcolortrue
  }{}%
  \@ifundefined{ifGPblacktext}{%
    \newif\ifGPblacktext
    \GPblacktexttrue
  }{}%
  \let\gplgaddtomacro\g@addto@macro
  \gdef\gplbacktext{}%
  \gdef\gplfronttext{}%
  \makeatother
  \ifGPblacktext
    \def\colorrgb#1{}%
    \def\colorgray#1{}%
  \else
    \ifGPcolor
      \def\colorrgb#1{\color[rgb]{#1}}%
      \def\colorgray#1{\color[gray]{#1}}%
      \expandafter\def\csname LTw\endcsname{\color{white}}%
      \expandafter\def\csname LTb\endcsname{\color{black}}%
      \expandafter\def\csname LTa\endcsname{\color{black}}%
      \expandafter\def\csname LT0\endcsname{\color[rgb]{1,0,0}}%
      \expandafter\def\csname LT1\endcsname{\color[rgb]{0,1,0}}%
      \expandafter\def\csname LT2\endcsname{\color[rgb]{0,0,1}}%
      \expandafter\def\csname LT3\endcsname{\color[rgb]{1,0,1}}%
      \expandafter\def\csname LT4\endcsname{\color[rgb]{0,1,1}}%
      \expandafter\def\csname LT5\endcsname{\color[rgb]{1,1,0}}%
      \expandafter\def\csname LT6\endcsname{\color[rgb]{0,0,0}}%
      \expandafter\def\csname LT7\endcsname{\color[rgb]{1,0.3,0}}%
      \expandafter\def\csname LT8\endcsname{\color[rgb]{0.5,0.5,0.5}}%
    \else
      \def\colorrgb#1{\color{black}}%
      \def\colorgray#1{\color[gray]{#1}}%
      \expandafter\def\csname LTw\endcsname{\color{white}}%
      \expandafter\def\csname LTb\endcsname{\color{black}}%
      \expandafter\def\csname LTa\endcsname{\color{black}}%
      \expandafter\def\csname LT0\endcsname{\color{black}}%
      \expandafter\def\csname LT1\endcsname{\color{black}}%
      \expandafter\def\csname LT2\endcsname{\color{black}}%
      \expandafter\def\csname LT3\endcsname{\color{black}}%
      \expandafter\def\csname LT4\endcsname{\color{black}}%
      \expandafter\def\csname LT5\endcsname{\color{black}}%
      \expandafter\def\csname LT6\endcsname{\color{black}}%
      \expandafter\def\csname LT7\endcsname{\color{black}}%
      \expandafter\def\csname LT8\endcsname{\color{black}}%
    \fi
  \fi
  \setlength{\unitlength}{0.0500bp}%
  \begin{picture}(4392.00,3074.40)%
    \gplgaddtomacro\gplbacktext{%
      \csname LTb\endcsname%
      \put(672,512){\makebox(0,0)[r]{\strut{} 0}}%
      \put(672,761){\makebox(0,0)[r]{\strut{} 0.1}}%
      \put(672,1011){\makebox(0,0)[r]{\strut{} 0.2}}%
      \put(672,1260){\makebox(0,0)[r]{\strut{} 0.3}}%
      \put(672,1510){\makebox(0,0)[r]{\strut{} 0.4}}%
      \put(672,1759){\makebox(0,0)[r]{\strut{} 0.5}}%
      \put(672,2009){\makebox(0,0)[r]{\strut{} 0.6}}%
      \put(672,2258){\makebox(0,0)[r]{\strut{} 0.7}}%
      \put(672,2508){\makebox(0,0)[r]{\strut{} 0.8}}%
      \put(672,2757){\makebox(0,0)[r]{\strut{} 0.9}}%
      \put(768,352){\makebox(0,0){\strut{} 0}}%
      \put(1374,352){\makebox(0,0){\strut{} 0.5}}%
      \put(1981,352){\makebox(0,0){\strut{} 1}}%
      \put(2587,352){\makebox(0,0){\strut{} 1.5}}%
      \put(3193,352){\makebox(0,0){\strut{} 2}}%
      \put(3800,352){\makebox(0,0){\strut{} 2.5}}%
      \put(208,1697){\rotatebox{-270}{\makebox(0,0){\strut{}$1/\sigma ~ d\sigma/d|\eta|_{\ell}^{max}$}}}%
      \put(2435,112){\makebox(0,0){\strut{}$|\eta|_{\ell}^{max}$}}%
    }%
    \gplgaddtomacro\gplfronttext{%
      \csname LTb\endcsname%
      \put(2443,2752){\makebox(0,0)[r]{\strut{}\tiny Standard Model}}%
      \csname LTb\endcsname%
      \put(2443,2592){\makebox(0,0)[r]{\strut{}\tiny$f_{M2}/ \Lambda^{4}= 800$ TeV$^{-4}$}}%
      \csname LTb\endcsname%
      \put(2443,2432){\makebox(0,0)[r]{\strut{}\tiny$f_{T1}/ \Lambda^{4}= 25$ TeV$^{-4}$}}%
      \csname LTb\endcsname%
      \put(2443,2272){\makebox(0,0)[r]{\strut{}\tiny$f_{WWW}/ \Lambda^{2}= 10$ TeV$^{-2}$}}%
    }%
    \gplbacktext
    \put(0,0){\includegraphics{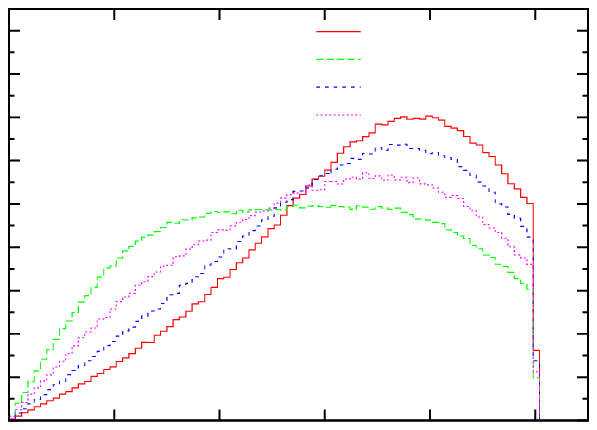}}%
    \gplfronttext
  \end{picture}%
\endgroup